\documentclass[prl,superscriptaddress,twocolumn,showpacs,preprintnumbers,amsmath,amssymb]{revtex4}

\usepackage{graphicx}
\usepackage{dcolumn}
\usepackage{bm}

\begin{document}

\title{Energy loss rates of hot Dirac fermions in epitaxial, exfoliated and CVD graphene}
\author{A.M.R. Baker}
\affiliation{Dept. of
Physics, University of Oxford, Clarendon Laboratory, Parks Rd.,
Oxford, OX1 3PU, U.K.}

\author{ J.A. Alexander-Webber}
\affiliation{Dept. of
Physics, University of Oxford, Clarendon Laboratory, Parks Rd.,
Oxford, OX1 3PU, U.K.}

\author{T. Altebaeumer}
\affiliation{Dept. of
Physics, University of Oxford, Clarendon Laboratory, Parks Rd.,
Oxford, OX1 3PU, U.K.}

\author{S.D. McMullan}
\affiliation{Dept. of
Physics, University of Oxford, Clarendon Laboratory, Parks Rd.,
Oxford, OX1 3PU, U.K.}

\author{T.J.B.M. Janssen}
\affiliation{National Physical Laboratory, Hampton Road, Teddington, TW11 0LW, U.K.}

\author{A. Tzalenchuk}
\affiliation{National Physical Laboratory, Hampton Road, Teddington, TW11 0LW, U.K.}

\author{S. Lara-Avila}
\affiliation{Dept. of Microtechnology and Nanoscience, Chalmers University of Technology, S-412 96 G$\ddot{o}$teborg, Sweden}

\author{S. Kubatkin}
\affiliation{Dept. of Microtechnology and Nanoscience, Chalmers University of Technology, S-412 96 G$\ddot{o}$teborg, Sweden}

\author{R. Yakimova}
\affiliation{Dept. of Physics, Chemistry and Biology, Link$\ddot{o}$ping University, S-581 83 Link$\ddot{o}$ping, Sweden}

\author{C.-T. Lin}
\affiliation{Institute of Atomic and Molecular Sciences, Academia Sinica, Taipei, 11617, Taiwan}

\author{L.-J. Li}
\affiliation{Institute of Atomic and Molecular Sciences, Academia Sinica, Taipei, 11617, Taiwan}

\author{R.J. Nicholas}
\email{r.nicholas1@physics.ox.ac.uk}
\affiliation{Dept. of
Physics, University of Oxford, Clarendon Laboratory, Parks Rd.,
Oxford, OX1 3PU, U.K.}

\date{\today}
\begin{abstract}

Energy loss rates for hot carriers in graphene have been measured using graphene produced by epitaxial growth on SiC, exfoliation and chemical vapour deposition (CVD). It is shown that the temperature dependence of the energy loss rates measured with high-field damped Shubnikov-de Haas oscillations, and the temperature dependence of the weak localization peak close to zero field correlate well, with the high-field measurements understating the energy loss rates by $\sim$40\% compared to the low-field results. The energy loss rates for all graphene samples follow a universal scaling of $T_{e}^4$ at low temperatures and depend weakly on carrier density $\propto$ n$^{-\frac{1}{2}}$ evidence for enhancement of the energy loss rate due to disorder in CVD samples. 


\end{abstract}

\pacs{73.43.Qt, 72.80.Vp, 72.10.Di}

\maketitle

\section{Introduction}

Despite its remarkable properties, a key problem in the commercialization of graphene has been that the fabrication method used for its discovery, micromechanical exfoliation\cite{Novoselov2004}, is not amenable to large-scale commercial production. There are, however, other production methods which can easily be scaled up, such as epitaxial growth on SiC, or chemical vapour deposition (CVD) on thin metal films\cite{vanNoorden2012}. While both these techniques are well established, the majority of graphene research is still carried out on exfoliated graphene. The principal reasons for this are the comparative ease of graphene fabrication using this method, and that to date it still produces the highest quality samples\cite{Bolotin2008}. There is therefore a great need for comparisons to be made between graphene produced by the `research' method, and those produced in a more commercially amenable manner.

Carrier energy loss rates are a particularly important parameter as they influence thermal dissipation and heat management in modern electronics as well as low-temperature applications such as quantum resistance metrology\cite{Tzalenchuk2010} and hot-electron bolometers\cite{Yan2012}. Energy loss rates have previously been measured in exfoliated graphene\cite{Baker2012, Tan2011,Betz2012}, with conflicting results. There has also been theoretical disagreement as to how Dirac fermions in graphene lose energy to the lattice and how this varies with temperature and carrier density\cite{Kubakaddi2009, Tse2009}. Here we compare the temperature dependence of the carrier energy loss rates in graphene produced by three different fabrication methods: exfoliation, CVD and epitaxial growth on SiC, for carrier densities ranging from 1 x 10$^{11}$\,cm$^{-2}$ to 1.6 x 10$^{13}$\,cm$^{-2}$ using two independent methods (weak localization (WL) and Shubnikov-de Haas (SdH) oscillations) and demonstrate that a single consistent picture exists.

\begin{figure}
\begin{center}
\includegraphics[width = 8.5cm]{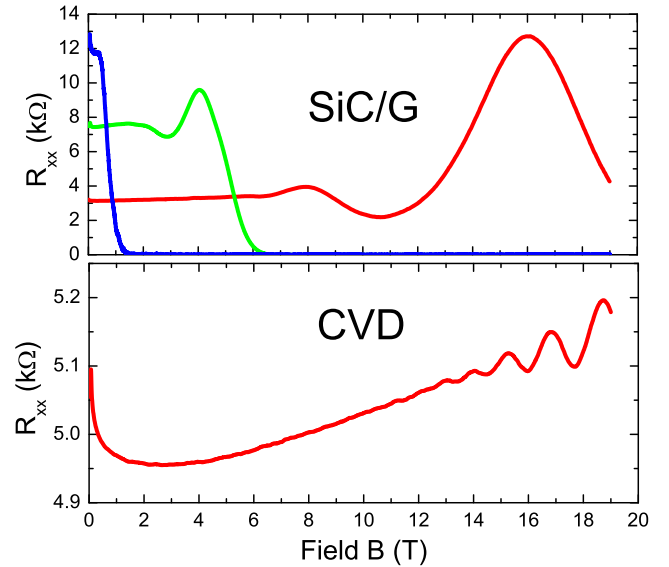}
\end{center}
\caption{Example $R_{xx}$ for the SiC/G and CVD samples. SiC/G, graphene grown epitaxially on SiC, before (Red, n = 1.63 x 10$^{12}$cm$^{-2}$, 160$\mu$m x 35$\mu$m),  after (Green, n = 4.72 x 10$^{11}$cm$^{-2}$), and after further (Blue, n = 1 x 10$^{11}$cm$^{-2}$) photochemical gating. CVD, graphene grown by CVD onto thin film metal, (n = 1.62 x 10$^{13}$cm$^{-2}$, 120$\mu$m x 32$\mu$m). } \label{fig:ExampleTraces}
\end{figure}

\section{Methodology}

\subsection{Samples}

The CVD graphene was grown on thin-film copper, subsequently transferred to Si/SiO$_2$\cite{Chen2012}. The graphene wafers were lithographed by e-beam into Hall-bars using gold-only final contacting as described in our previous work on exfoliated graphene\cite{Baker2012}. The SiC/Graphene (SiC/G) was epitaxially grown on the Si-terminated face of SiC\cite{Janssen2011}. Hall-bar devices were produced using e-beam lithography and oxygen plasma etching with large area titanium-gold contacting. UV exposure of top surface polymers was used to control the carrier density \cite{Lara2011}. The carrier mobilities as deduced from the sample resisitivity and the two dimensional carrier density were strongly carrier density dependent, falling from 24,000 cm$^2$V$^{-1}$s$^{-1}$ at n = 1 x 10$^{11}$ cm$^{-2}$ to 6,000 cm$^2$V$^{-1}$s$^{-1}$ at n = 1.6 x 10$^{12}$ cm$^{-2}$ for SiC/G and to 300 cm$^2$V$^{-1}$s$^{-1}$ at n = 1.6 x 10$^{13}$ cm$^{-2}$ for the CVD graphene.  All electrical measurements were carried out using DC constant-current sources and multimeters.

\section{Results}

\subsection{Shubnikov-deHaas oscillations}

The conventional method of measuring the temperature dependence of the energy loss rates is by comparison of the temperature-induced damping of SdH oscillations with that induced by current heating. In both cases the damping is caused by thermal broadening of the carrier distribution. This method can be used for high density samples with a sufficient mobility to display sinusoidal oscillations. Using the well known Lifshitz-Kosevich formula, which has been shown to apply to both conventional\cite{Ando1982} and Dirac-like\cite{Sharapov2004} two dimensional systems, we calculate the carrier temperature ($T_{e}$) from the damped amplitude with

\begin{equation}\label{eqn:TempAmp}
\frac{\Delta \rho}{\rho} = f(\omega_c \tau)\frac{\chi}{\sinh\chi}e^{-\frac{\pi}{\omega_c\tau_q}},\end{equation}

\noindent where $\tau_q$ is the quantum lifetime and

\begin{equation}\label{eqn:MyChi}
\chi = \frac{2 \pi^2k_{B}T_e} {\hbar \omega_c},
\end{equation}

\noindent and $\hbar \omega_c$ is calculated from the separation of the N and N+1 Landau levels from

\begin{equation}
  E_{N} = \textrm{sgn(N)}\times v_{F} \sqrt{2e\hbar B|N|},
  \label{eqn"LL}
  \end{equation}

\noindent where B is the magnetic field and $v_F$ is the Fermi velocity, which is 1.1 x 10$^6$ms$^{-1}$ as measured in both epitaxial SiC/G\cite{Miller2009} and exfoliated material \cite{Deacon2007}.

\begin{table} [b]
\footnotesize
\begin{center}
\begin{tabular} {cccc}
\hline
Filling&Field&Landau&Velocity\\
factor $\nu$&B(T)& level N&v$_F$ (ms$^{-1}$)\\
\hline
34&17.69&8&1.06 x 10$^6$\\
38&15.83&9&1.09 x 10$^6$\\
42&14.33&10&1.10 x 10$^6$\\
\hline
&&Average v$_F$ =& 1.083 x 10$^6$\\
\hline
\end{tabular}
\end{center}
\caption{Values of v$_F$ calculated from the fitting of the temperature dependence of the Shubnikov-de Haas amplitudes, taken in CVD graphene with a carrier density of 1.43 x 10$^{13}$ cm$^{-2}$.} \label{fig:vTable}
\end{table}

\begin{figure}
\begin{center}
\includegraphics[width = 9cm]{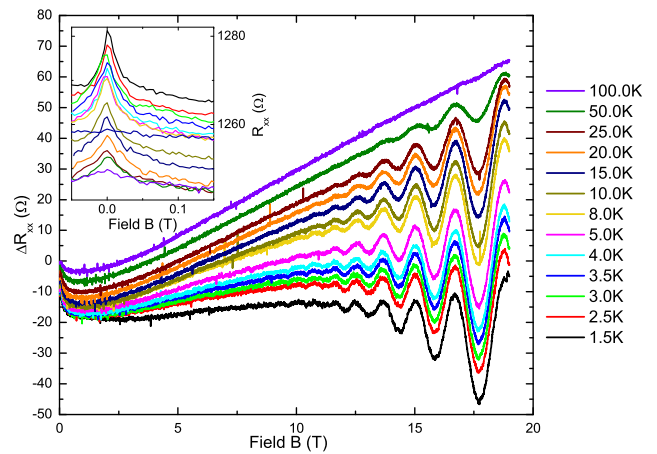}
\end{center}
\caption{R$_{xx}$ magnetic field traces taken in CVD graphene with n = 1.43 x 10$^{13}$ cm$^{-2}$. All traces were taken at 500nA from 1.5K to 100K. Traces are shifted such that the peak of the weak-localization for all traces lies at 0$\Omega$. The inset expands the region around B = 0T to show clearly the weak-localization peaks.} \label{fig:CVDFullTemps}
\end{figure} 

\begin{figure}
\begin{center}
\includegraphics[width = 9cm]{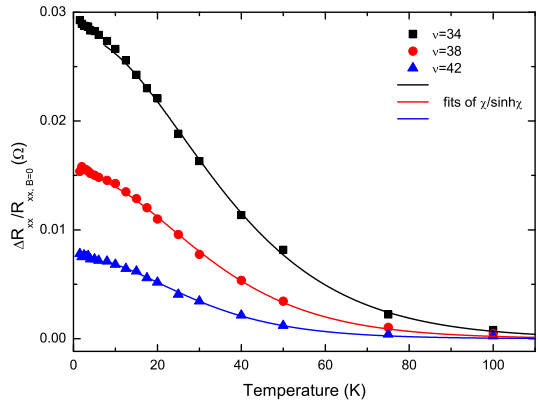}
\end{center}
\caption{Temperature dependence of the normalized amplitude of Shubnikov-de Haas oscillations from a CVD sample with n = 1.43 x 10$^{13}$ cm$^{-2}$, at the three smallest measured filling factors. Fits of $\frac{\chi}{\sinh\chi}$ are shown from which the Fermi velocity is derived.} \label{fig:CVDv}
\end{figure}

For the CVD material the carrier densities are nearly an order of magnitude higher than studied previously. To validate the method at these high carrier densities we measured the lattice temperature dependence of the damping, as shown in Fig. \ref{fig:CVDFullTemps}, using a sufficiently low current to avoid heating the carriers above the lattice temperature. Fig. \ref{fig:CVDv} shows these amplitudes as a function of temperature fitted using Eq.\ref{eqn:TempAmp}. The data are well fitted by this equation, and the Fermi velocity, v$_F$, is deduced from Eq. \ref{eqn"LL}. The measured values of v$_F$ are shown in Table \ref{fig:vTable} for a carrier density of 1.43 x 10$^{13}$ cm$^{-2}$. The mean value of v$_F$ = 1.083 x 10$^6$ $\pm$ 0.02 x 10$^6$ ms$^{-1}$ is consistent with the values reported for exfoliated and epitaxial graphene at lower densities which shows that the Fermi velocity remains essentially constant up to energies of at least E$_F$ = 480meV.

\begin{figure}
\begin{center}
\includegraphics[width = 8.5cm]{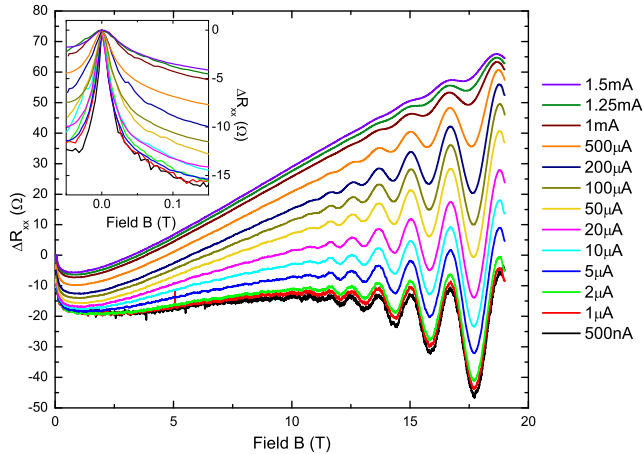}
\end{center}
\caption{R$_{xx}$ for CVD graphene with n = 1.43 x 10$^{13}$\,cm$^{-2}$. All traces taken at 1.5K with the current varied from 500nA up to 1.5mA. Traces are shifted such that the WL peak is at 0$\Omega$. The inset shows the WL peak around B = 0T.} \label{fig:CVDFullCurrents}
\end{figure}

When the ambient temperature is fixed, but a current is passed through the sample, the carriers are unable to lose energy at a sufficient rate to reach thermal equilibrium with the lattice and thus heat up\cite{Leadley1989,Leturcq2003,Fletcher1992,Gershenson2000,Stoger1994b}. The energy loss rates were determined as a function of carrier temperature by measuring the amplitudes of the SdH oscillations as a function of current (Fig.\ref{fig:CVDFullCurrents}). Comparing the amplitudes to the undamped value, taken at low current, and using Eq.(\ref{eqn:TempAmp}), the carrier temperature for each trace is determined. The associated energy loss rate per carrier (P) is calculated using

\begin{equation}\label{eqn:ELR}
P = \frac{I^2 R_{xx}}{nA},
\end{equation}

where $R_{xx}$ is the longitudinal sample resistance, n is the carrier density, and A is the device area between the resistivity contacts where the power is dissipated.

\subsection{Weak Localization}

For some samples, however, the SdH method for extracting energy loss rates could not be applied, most notably for the lower carrier density SiC/G devices, where the carrier densities are too low to support a series of sinusoidal SdH oscillations (Fig. \ref{fig:ExampleTraces}). SiC/G samples do, however, have prominent WL peaks.

The magnitude of the WL peak is known\cite{Cao2010,Lara2011b,Baker2012b} to be strongly damped by increasing temperature. Energy loss rates have previously been measured as a function of carrier temperature from WL peaks for several other materials\cite{Leturcq2003,Fletcher1992,Gershenson2000,Stoger1994b}. In GaAs\cite{Fletcher1992} good agreement was found between the values for electron temperature deduced from the WL peak and SdH oscillations at low fields ($\sim$0.5T) up to $\sim$4K. Similarly excellent agreement was found for the two methods in SiGe\cite{Leturcq2003} but could only be studied for electron temperatures up to 1.1K. In our work the WL peak persists up to temperatures of almost 100K allowing a much more extensive range of parameters to be studied. Previous work in graphene\cite{Lee2010}, however, was not able to determine energy loss rates from the damping of the WL peak.

\begin{figure}
\begin{center}
\includegraphics[width = 8.5cm]{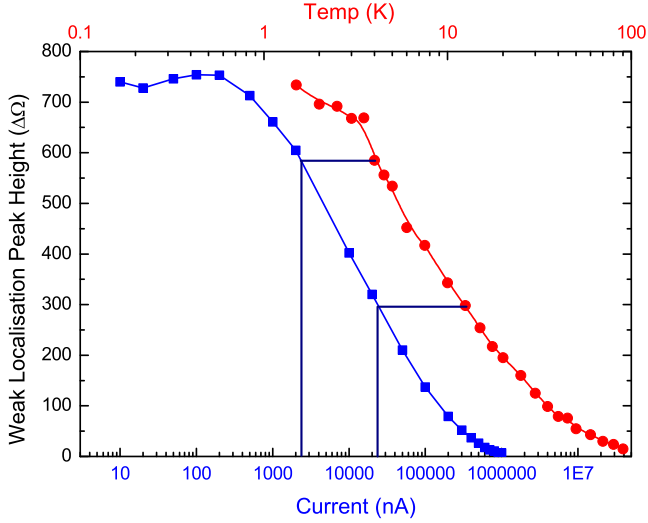}
\end{center}
\caption{An example set of data displaying the matching of the height of the WL peak as a function of current (blue squares) and temperature (red circles), from a SiC sample with n = 4.13 x 10$^{11}$\,cm$^{-2}$.} \label{fig:TempCurrentComparison}
\end{figure}

We determine the energy loss rates by comparison of the current and temperature dependence of the WL peak. The magnitudes of the peaks are measured by taking the difference of R$_{xx}$ between 0T and a fixed small field, sufficient to entirely suppress the WL behavior.  The temperature dependent peak height measured at low current is fitted with an interpolated curve. The magnitude of the WL peak measured as a function of current is then compared with the interpolated temperature dependence and used to deduce the carrier temperature as a function of current (Fig. \ref{fig:TempCurrentComparison}).

It is worth emphasizing that for the above analysis to be correct, the effects of raising both the carrier and lattice temperature (T$_L$) need to be equivalent. For this to be the case the inelastic scattering time $\tau_\varphi$ must only depend on the carrier temperature, and be independent of the lattice temperature. This has been verified by Ki \textit{et al.}\cite{Ki2008} by measuring the effect on $\tau_\varphi$ of changes to the carrier density at temperatures up to 20K. The inelastic scattering, $\tau_\varphi$, in graphene was shown to depend on Coulomb interactions and Nyquist scattering, both of which vary with carrier temperature only. This can be qualitatively seen by the similar functional dependence upon current and temperature, shown in Fig. \ref{fig:TempCurrentComparison}.

\subsection{Energy Loss Rates}

\begin{figure}
\begin{center}
\includegraphics[width = 8.5cm]{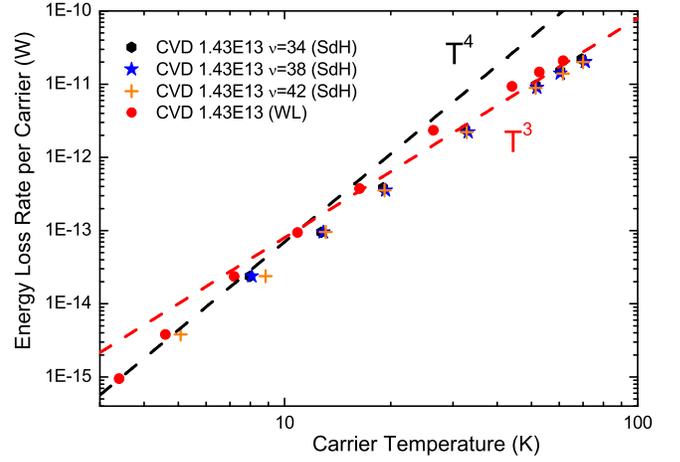}
\end{center}
\caption{Carrier energy loss rate as a function of electron temperature for a CVD sample as measured by both the WL and the SdH technique at occupancies of $\nu$=34, 38 and 42. Power law dependences are shown for T$^4$ and T$^3$ for comparison.} \label{fig:CVDcomp}
\end{figure}

The temperature dependence of the energy loss rates as measured by both techniques up to 90K are shown in Fig. \ref{fig:CVDcomp} for a CVD sample which compares the results for the WL method and analysis using the SdH method at three different occupancies.  This demonstrates that there is a power law dependence of the energy loss rates which tends towards T$^{4}$ at low temperature and that there is also a systematic difference between the two methods with the SdH technique giving energy loss rates $\sim$40\% lower (Table \ref{fig:Table}) at any given electron temperature. This difference is in contrast to GaAs\cite{Fletcher1992} and SiGe\cite{Leturcq2003} where good agreement between the two methods was found when SdH oscillations below 1T were analysed. We attribute this to a magnetic-field suppression of the energy loss rate, as the WL method measures the energy loss rate near 0T, whereas the relatively low mobilities in most graphene samples require the SdH method to make measurements at much higher fields, typically 10T to 18T in our samples, where the Landau quantization energy is 30 to 50 meV and significant changes in the current distribution also occur. No systematic dependence of electron temperature as a function of occupancy factor, $\nu$, could however, be detected at high fields, as shown in Fig. \ref{fig:CVDcomp} for $\nu$=34, 38 and 42. Taken altogether therefore, these methods allow the systematic measurement of the energy loss rates per carrier as a function of carrier temperature for any graphene sample, and permit a robust comparison to be made between such samples.

\begin{table}
\footnotesize
\begin{center}

\begin{tabular} {ccccc}
\hline
Sample Type & Carrier Density & SdH $\alpha$& WL $\alpha$&$\alpha$SdH /\\
&(cm$^{-2}$) &(W\,K$^{-4}$/\,&(W\,K$^{-4}$/\,&$\alpha$WL\\
& & carrier)& carrier)\\
\hline
CVD & $1.62 \cdot 10^{13}$ & 2.4 $\cdot 10^{-18}$& 3.8 $\cdot 10^{-18}$ & 0.63\\
CVD & $1.43 \cdot 10^{13}$ & 2.0 $\cdot 10^{-18}$ & 3.0 $\cdot 10^{-18}$ & 0.67\\
CVD & $8.16 \cdot 10^{12}$ & 2.2 $\cdot 10^{-18}$ &3.7 $\cdot 10^{-18}$ & 0.59\\
Epitaxial & $1.63 \cdot 10^{12}$ & 2.1 $\cdot 10^{-18}$ &3.7 $\cdot 10^{-18}$ & 0.57\\
\hline
&&&Average = &0.62\\
\hline
\end{tabular}
\end{center}
\caption{The $\alpha$ values $P = \alpha(T^4_e-T_L^4)$ for CVD and SiC/G samples measured by the SdH and WL methods.} \label{fig:Table}
\end{table}

\begin{figure}
\begin{center}
\includegraphics[width = 8.5cm]{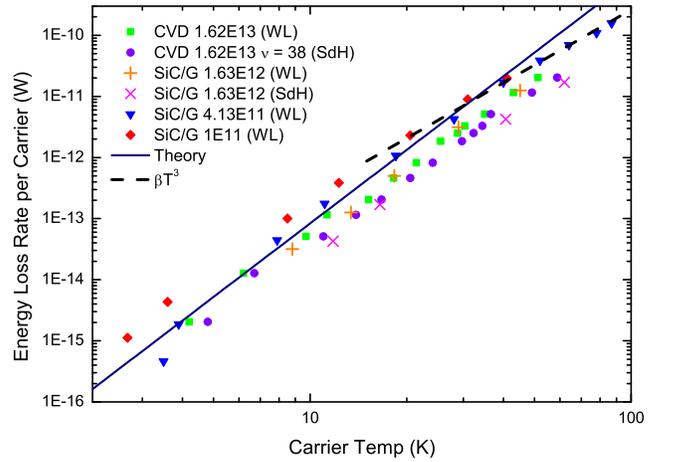}
\end{center}
\caption{Carrier energy loss rate as a function of electron temperature for a representative variety of samples and from the two techniques used. The data follow a similar trend for all samples, with the energy loss rate scaling by approximately T$^4$. An extrapolated T$^4$ dependence from low temperature calculations\cite{Kubakaddi2009} is shown for a carrier density of 4.13 x 10$^{11}$ cm$^{-2}$ and a T$^3$ dependence is also shown.} \label{fig:AllELRs}
\end{figure}

Fig. \ref{fig:AllELRs} summarises the results from both techniques on all of the types of graphene, with the exception that only the SdH method could be used on the exfoliated graphene, which does not exhibit large WL peaks\cite{morozov2006}. Overall this demonstrates that there are no significant differences in the systematic behaviour between samples derived from the different production methods. We find that samples produced by all the measured production methods are well fitted to a low temperature, limiting behaviour of $P = \alpha(n)(T^4_e-T_L^4)$, where $\alpha(n)$ is a scaling constant that is weakly carrier density dependent. This can be approximated to the form $P = \alpha(T^4_e)$ when T$_L$ $\ll$ T$_e$. Theoretical predictions of the energy loss rate at low temperatures from Kubakaddi\cite{Kubakaddi2009} are shown for the example of a carrier density of n = 4.13 x 10$^{11}$\,cm$^{-2}$. These are dominated by deformation potential coupling and are shown to be an excellent fit with a deformation potential of 19 eV. This $(T^4_e)$ power law behaviour is typical for low temperatures
below the Bloch-Gr\"{u}neisen temperature ($T_{BG}$) as observed for the resistivity in graphene\cite{EfKim2010} and is broadly consistent with the results of Betz \textit{et al.}\cite{Betz2012} who reported an approximate $(T^4_e)$ power law at high electron temperatures, but with a reduced coupling constant. This is in contrast to some previous theoretical\cite{Tse2009} and experimental\cite{Tan2011} work which suggested other powers of T.

The value of $T_{BG}$ (=$\hbar v_s k_F/k_{B})$, where $k_F$ is the Fermi wavevector) for the samples studied is in the range 10-125K (1-160 x 10$^{11}$\,cm$^{-2}$), but as found previously in GaAs\cite{Leadley1989} the data show no evidence of approaching a high temperature limit of linear T dependence, even at temperatures significantly above $T_{BG}$. At the highest temperatures there is however some decrease in the energy loss rate, closer to a power law in the region of $T^{3}$ or lower, as reported previously for exfoliated graphene\cite{Baker2012} at high $T_e$. It is likely that at higher temperatures above 100K some additional contribution may occur due to optical phonons in both the graphene and the substrate\cite{Tse2009}\cite{newChen08}.

\begin{figure}
\begin{center}
\includegraphics[width = 8.3cm]{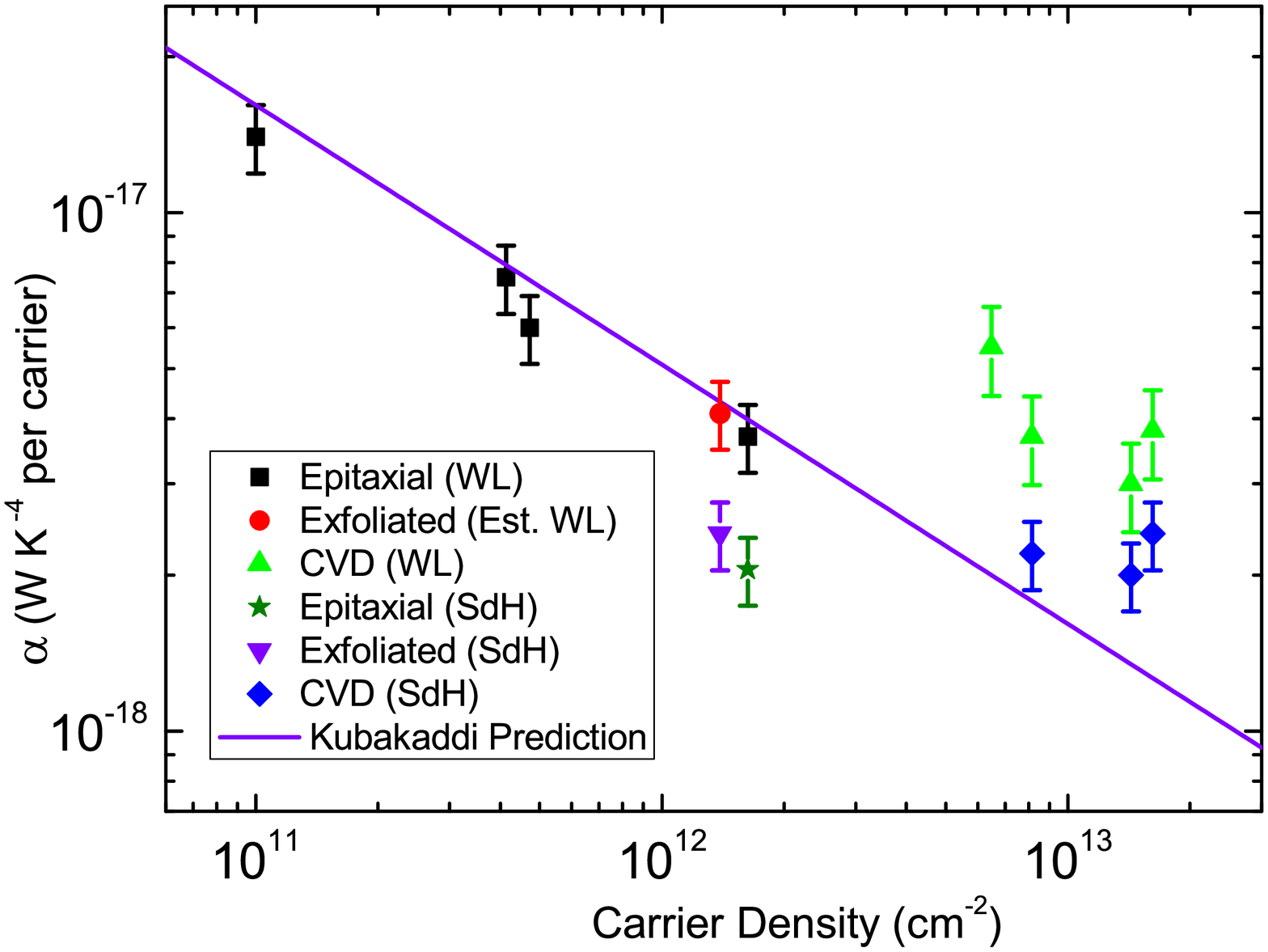}
\end{center}
\caption{The $\alpha$ values as a function of carrier density for the CVD and epitaxial graphene data deduced from WL, and SdH methods. The point labelled in the key as exfoliated (Est. WL)is estimated from previous SdH results for an exfoliated sample\cite{Baker2012} by adjusting the value upwards by a factor 1.6 to take account of the the measured systematic difference between the two methods(Table \ref{fig:Table}). Solid line on the graph is the theoretical prediction from Kubakaddi\cite{Kubakaddi2009} using a deformation-potential coupling constant of 19 eV.} \label{fig:Trend} 
\end{figure} 

Fig. \ref{fig:Trend} shows the fitted $\alpha$ coefficient for data below 30K for all the measured samples, and where possible for multiple measurement techniques, plotted against carrier density. The theoretical predictions\cite{Kubakaddi2009} of the absolute values are in excellent agreement with experiment for the epitaxial and exfoliated samples and show an $\alpha \propto n^{-\frac{1}{2}}$ dependence on carrier density as predicted. However for CVD samples the energy loss rates appear to be a factor of 2 - 3 larger. Recent theoretical work from Song et al.\cite{Song2012} has proposed enhanced energy loss processes in graphene which they term ``supercollisions''. These are caused by disorder-assisted, and two-phonon scattering at high k values and are particularly important for higher energy carriers which play a disproportionately large role in the energy loss processes. Our CVD samples contain significant amounts of disorder and hence the ``supercollision'' process could potentially account for the increased energy loss rates of the CVD samples relative to the theoretical predictions\cite{Kubakaddi2009}. Song et al. predict that for strongly disordered samples (using $k_{F}l \sim$ 2) this should give a dependence $\propto T^{3}$ above $T_{BG}$, which has recently received experimental support from measurements of optical heating\cite{Graham2013} and noise thermometry\cite{Betz2013}. Fitting the higher temperature data for the CVD samples in Fig. \ref{fig:AllELRs} gives P $\propto T^{3}$, suggesting that this process is beginning to contribute to the energy loss rate.

\section{Conclusions}

In conclusion we have performed an extensive comparison of the transport properties of graphene produced by the three most common production methods. The energy loss rates per carrier as a function of carrier temperature follow a low temperature limiting T$^4$ dependence in all cases, with a multiplicative factor that varies weakly with carrier density in good agreement with theory\cite{Kubakaddi2009} based on deformation potential scattering. The close agreement between epitaxial graphene grown on SiC and exfoliated flakes deposited on SiO$_{2}$ suggest that the substrate plays very little role in determining the energy loss rates. An enhanced energy loss rate is observed in CVD samples deposited onto SiO$_{2}$ which may be attributable to ``supercollisions''\cite{Song2012} allowed by the presence of increased disorder and which may be beneficial for high power electronic applications of graphene. We have also established that energy loss rates can be measured over a wide temperature range using the WL correspondence method and correlate well with the conventional SdH method, but the values deduced at high fields (10-18T) are suppressed by $\sim$40\% relative to the low field values.

\section{Acknowledgements}
This work was supported by the UK EPSRC, Swedish Research Council and Foundation for Strategic Research, UK National Measurement Office, and EU FP7 STREP ConceptGraphene.


\begin{thebibliography}{33}
\expandafter\ifx\csname natexlab\endcsname\relax\def\natexlab#1{#1}\fi
\expandafter\ifx\csname bibnamefont\endcsname\relax
  \def\bibnamefont#1{#1}\fi
\expandafter\ifx\csname bibfnamefont\endcsname\relax
  \def\bibfnamefont#1{#1}\fi
\expandafter\ifx\csname citenamefont\endcsname\relax
  \def\citenamefont#1{#1}\fi
\expandafter\ifx\csname url\endcsname\relax
  \def\url#1{\texttt{#1}}\fi
\expandafter\ifx\csname urlprefix\endcsname\relax\def\urlprefix{URL }\fi
\providecommand{\bibinfo}[2]{#2}
\providecommand{\eprint}[2][]{\url{#2}}

\bibitem[{\citenamefont{Novoselov et~al.}(2004)\citenamefont{Novoselov, Geim,
  Morozov, Jiang, Zhang, Dubonos, Grigorieva, and Firsov}}]{Novoselov2004}
\bibinfo{author}{\bibfnamefont{K.~S.} \bibnamefont{Novoselov}},
  \bibinfo{author}{\bibfnamefont{A.~K.} \bibnamefont{Geim}},
  \bibinfo{author}{\bibfnamefont{S.~V.} \bibnamefont{Morozov}},
  \bibinfo{author}{\bibfnamefont{D.}~\bibnamefont{Jiang}},
  \bibinfo{author}{\bibfnamefont{Y.}~\bibnamefont{Zhang}},
  \bibinfo{author}{\bibfnamefont{S.~V.} \bibnamefont{Dubonos}},
  \bibinfo{author}{\bibfnamefont{I.~V.} \bibnamefont{Grigorieva}},
  \bibnamefont{and} \bibinfo{author}{\bibfnamefont{A.~A.}
  \bibnamefont{Firsov}}, \bibinfo{journal}{Science}
  \textbf{\bibinfo{volume}{306}}, \bibinfo{pages}{666} (\bibinfo{year}{2004}).

\bibitem[{\citenamefont{van Noorden}(2012)}]{vanNoorden2012}
\bibinfo{author}{\bibfnamefont{R.}~\bibnamefont{van Noorden}},
  \bibinfo{journal}{Nature} \textbf{\bibinfo{volume}{483}},
  \bibinfo{pages}{S32} (\bibinfo{year}{2012}).

\bibitem[{\citenamefont{Bolotin et~al.}(2008)\citenamefont{Bolotin, Sikes,
  Jiang, Klima, Fudenberg, Hone, Kim, and Stormer}}]{Bolotin2008}
\bibinfo{author}{\bibfnamefont{K.}~\bibnamefont{Bolotin}},
  \bibinfo{author}{\bibfnamefont{K.}~\bibnamefont{Sikes}},
  \bibinfo{author}{\bibfnamefont{Z.}~\bibnamefont{Jiang}},
  \bibinfo{author}{\bibfnamefont{M.}~\bibnamefont{Klima}},
  \bibinfo{author}{\bibfnamefont{G.}~\bibnamefont{Fudenberg}},
  \bibinfo{author}{\bibfnamefont{J.}~\bibnamefont{Hone}},
  \bibinfo{author}{\bibfnamefont{P.}~\bibnamefont{Kim}}, \bibnamefont{and}
  \bibinfo{author}{\bibfnamefont{H.}~\bibnamefont{Stormer}},
  \bibinfo{journal}{Solid State Commun.} \textbf{\bibinfo{volume}{146}},
  \bibinfo{pages}{351} (\bibinfo{year}{2008}).

\bibitem[{\citenamefont{Tzalenchuk et~al.}(2010)\citenamefont{Tzalenchuk,
  Lara-Avila, Kalaboukhov, Paolillo, Syv\"{a}j\"{a}rvi, Yakimova, Kazakova,
  Janssen, Fal'ko, and Kubatkin}}]{Tzalenchuk2010}
\bibinfo{author}{\bibfnamefont{A.}~\bibnamefont{Tzalenchuk}},
  \bibinfo{author}{\bibfnamefont{S.}~\bibnamefont{Lara-Avila}},
  \bibinfo{author}{\bibfnamefont{A.}~\bibnamefont{Kalaboukhov}},
  \bibinfo{author}{\bibfnamefont{S.}~\bibnamefont{Paolillo}},
  \bibinfo{author}{\bibfnamefont{M.}~\bibnamefont{Syv\"{a}j\"{a}rvi}},
  \bibinfo{author}{\bibfnamefont{R.}~\bibnamefont{Yakimova}},
  \bibinfo{author}{\bibfnamefont{O.}~\bibnamefont{Kazakova}},
  \bibinfo{author}{\bibfnamefont{T.~J. B.~M.} \bibnamefont{Janssen}},
  \bibinfo{author}{\bibfnamefont{V.}~\bibnamefont{Fal'ko}}, \bibnamefont{and}
  \bibinfo{author}{\bibfnamefont{S.}~\bibnamefont{Kubatkin}},
  \bibinfo{journal}{Nat. Nanotechnol.} \textbf{\bibinfo{volume}{5}},
  \bibinfo{pages}{186} (\bibinfo{year}{2010}), ISSN \bibinfo{issn}{1748-3395}.

\bibitem[{\citenamefont{Yan et~al.}(2012)\citenamefont{Yan, Kim, Elle, Sushkov,
  Jenkins, Milchberg, Fuhrer, and Drew}}]{Yan2012}
\bibinfo{author}{\bibfnamefont{J.}~\bibnamefont{Yan}},
  \bibinfo{author}{\bibfnamefont{M.-H.} \bibnamefont{Kim}},
  \bibinfo{author}{\bibfnamefont{J.~A.} \bibnamefont{Elle}},
  \bibinfo{author}{\bibfnamefont{A.~B.} \bibnamefont{Sushkov}},
  \bibinfo{author}{\bibfnamefont{G.~S.} \bibnamefont{Jenkins}},
  \bibinfo{author}{\bibfnamefont{H.~M.} \bibnamefont{Milchberg}},
  \bibinfo{author}{\bibfnamefont{M.~S.} \bibnamefont{Fuhrer}},
  \bibnamefont{and} \bibinfo{author}{\bibfnamefont{H.~D.} \bibnamefont{Drew}},
  \bibinfo{journal}{Nat. Nanotechnol.} \textbf{\bibinfo{volume}{7}},
  \bibinfo{pages}{472} (\bibinfo{year}{2012}).

\bibitem[{\citenamefont{Baker et~al.}(2012{\natexlab{a}})\citenamefont{Baker,
  Alexander-Webber, Altebaeumer, and Nicholas}}]{Baker2012}
\bibinfo{author}{\bibfnamefont{A.~M.~R.} \bibnamefont{Baker}},
  \bibinfo{author}{\bibfnamefont{J.~A.} \bibnamefont{Alexander-Webber}},
  \bibinfo{author}{\bibfnamefont{T.}~\bibnamefont{Altebaeumer}},
  \bibnamefont{and} \bibinfo{author}{\bibfnamefont{R.~J.}
  \bibnamefont{Nicholas}}, \bibinfo{journal}{Physical Review B}
  \textbf{\bibinfo{volume}{85}}, \bibinfo{pages}{115403}
  (\bibinfo{year}{2012}{\natexlab{a}}).

\bibitem[{\citenamefont{Tan et~al.}(2011)\citenamefont{Tan, Tan, Ma, Liu, Lu,
  and Yang}}]{Tan2011}
\bibinfo{author}{\bibfnamefont{Z.}~\bibnamefont{Tan}},
  \bibinfo{author}{\bibfnamefont{C.}~\bibnamefont{Tan}},
  \bibinfo{author}{\bibfnamefont{L.}~\bibnamefont{Ma}},
  \bibinfo{author}{\bibfnamefont{G.~T.} \bibnamefont{Liu}},
  \bibinfo{author}{\bibfnamefont{L.}~\bibnamefont{Lu}}, \bibnamefont{and}
  \bibinfo{author}{\bibfnamefont{C.~L.} \bibnamefont{Yang}},
  \bibinfo{journal}{Phys. Rev. B.} \textbf{\bibinfo{volume}{84}},
  \bibinfo{pages}{115429} (\bibinfo{year}{2011}).

\bibitem[{\citenamefont{Betz et~al.}(2012)\citenamefont{Betz, Vialla, Brunel,
  Voisin, Picher, Cavanna, Madouri, Feve, Berroir, Placais et~al.}}]{Betz2012}
\bibinfo{author}{\bibfnamefont{A.}~\bibnamefont{Betz}},
  \bibinfo{author}{\bibfnamefont{F.}~\bibnamefont{Vialla}},
  \bibinfo{author}{\bibfnamefont{D.}~\bibnamefont{Brunel}},
  \bibinfo{author}{\bibfnamefont{C.}~\bibnamefont{Voisin}},
  \bibinfo{author}{\bibfnamefont{M.}~\bibnamefont{Picher}},
  \bibinfo{author}{\bibfnamefont{A.}~\bibnamefont{Cavanna}},
  \bibinfo{author}{\bibfnamefont{A.}~\bibnamefont{Madouri}},
  \bibinfo{author}{\bibfnamefont{G.}~\bibnamefont{Feve}},
  \bibinfo{author}{\bibfnamefont{J.-M.} \bibnamefont{Berroir}},
  \bibinfo{author}{\bibfnamefont{B.}~\bibnamefont{Placais}},
  \bibnamefont{et~al.}, \bibinfo{journal}{Phys. Rev. Lett.}
  \textbf{\bibinfo{volume}{109}}, \bibinfo{pages}{056805}
  (\bibinfo{year}{2012}).

\bibitem[{\citenamefont{Kubakaddi}(2009)}]{Kubakaddi2009}
\bibinfo{author}{\bibfnamefont{S.~S.} \bibnamefont{Kubakaddi}},
  \bibinfo{journal}{Physical Review B} \textbf{\bibinfo{volume}{79}},
  \bibinfo{pages}{075417} (\bibinfo{year}{2009}).

\bibitem[{\citenamefont{Tse and {Das Sarma}}(2009)}]{Tse2009}
\bibinfo{author}{\bibfnamefont{W.-K.} \bibnamefont{Tse}} \bibnamefont{and}
  \bibinfo{author}{\bibfnamefont{S.}~\bibnamefont{{Das Sarma}}},
  \bibinfo{journal}{Physical Review B} \textbf{\bibinfo{volume}{79}},
  \bibinfo{pages}{235406} (\bibinfo{year}{2009}).

\bibitem[{\citenamefont{Chen et~al.}(2012)\citenamefont{Chen, Lin, Lee, Liu,
  Su, Zhang, and Li}}]{Chen2012}
\bibinfo{author}{\bibfnamefont{C.-H.} \bibnamefont{Chen}},
  \bibinfo{author}{\bibfnamefont{C.-T.} \bibnamefont{Lin}},
  \bibinfo{author}{\bibfnamefont{Y.-H.} \bibnamefont{Lee}},
  \bibinfo{author}{\bibfnamefont{K.-K.} \bibnamefont{Liu}},
  \bibinfo{author}{\bibfnamefont{C.-Y.} \bibnamefont{Su}},
  \bibinfo{author}{\bibfnamefont{W.}~\bibnamefont{Zhang}}, \bibnamefont{and}
  \bibinfo{author}{\bibfnamefont{L.-J.} \bibnamefont{Li}},
  \bibinfo{journal}{Small} \textbf{\bibinfo{volume}{8}}, \bibinfo{pages}{43}
  (\bibinfo{year}{2012}), ISSN \bibinfo{issn}{1613-6829}.

\bibitem[{\citenamefont{Janssen et~al.}(2011)\citenamefont{Janssen, Tzalenchuk,
  Yakimova, Kubatkin, Lara-Avila, Kopylov, and Fal'ko}}]{Janssen2011}
\bibinfo{author}{\bibfnamefont{T.~J. B.~M.} \bibnamefont{Janssen}},
  \bibinfo{author}{\bibfnamefont{A.}~\bibnamefont{Tzalenchuk}},
  \bibinfo{author}{\bibfnamefont{R.}~\bibnamefont{Yakimova}},
  \bibinfo{author}{\bibfnamefont{S.}~\bibnamefont{Kubatkin}},
  \bibinfo{author}{\bibfnamefont{S.}~\bibnamefont{Lara-Avila}},
  \bibinfo{author}{\bibfnamefont{S.}~\bibnamefont{Kopylov}}, \bibnamefont{and}
  \bibinfo{author}{\bibfnamefont{V.~I.} \bibnamefont{Fal'ko}},
  \bibinfo{journal}{Phys. Rev. B.} \textbf{\bibinfo{volume}{83}},
  \bibinfo{pages}{233402} (\bibinfo{year}{2011}).

\bibitem[{\citenamefont{Lara-Avila
  et~al.}(2011{\natexlab{a}})\citenamefont{Lara-Avila, Moth-Poulsen, Yakimova,
  Bj\o~rnholm, Fal'ko, Tzalenchuk, and Kubatkin}}]{Lara2011}
\bibinfo{author}{\bibfnamefont{S.}~\bibnamefont{Lara-Avila}},
  \bibinfo{author}{\bibfnamefont{K.}~\bibnamefont{Moth-Poulsen}},
  \bibinfo{author}{\bibfnamefont{R.}~\bibnamefont{Yakimova}},
  \bibinfo{author}{\bibfnamefont{T.}~\bibnamefont{Bj\o~rnholm}},
  \bibinfo{author}{\bibfnamefont{V.}~\bibnamefont{Fal'ko}},
  \bibinfo{author}{\bibfnamefont{A.}~\bibnamefont{Tzalenchuk}},
  \bibnamefont{and} \bibinfo{author}{\bibfnamefont{S.}~\bibnamefont{Kubatkin}},
  \bibinfo{journal}{Adv. Mater.} \textbf{\bibinfo{volume}{23}},
  \bibinfo{pages}{878} (\bibinfo{year}{2011}{\natexlab{a}}).

\bibitem[{\citenamefont{Ando et~al.}(1982)\citenamefont{Ando, Fowler, and
  Stern}}]{Ando1982}
\bibinfo{author}{\bibfnamefont{T.}~\bibnamefont{Ando}},
  \bibinfo{author}{\bibfnamefont{A.~B.} \bibnamefont{Fowler}},
  \bibnamefont{and} \bibinfo{author}{\bibfnamefont{F.}~\bibnamefont{Stern}},
  \bibinfo{journal}{Rev. Mod. Physics} \textbf{\bibinfo{volume}{54}},
  \bibinfo{pages}{437} (\bibinfo{year}{1982}).

\bibitem[{\citenamefont{Gusynin and Sharapov}(2005)}]{Sharapov2004}
\bibinfo{author}{\bibfnamefont{V.~P.} \bibnamefont{Gusynin}} \bibnamefont{and}
  \bibinfo{author}{\bibfnamefont{S.~G.} \bibnamefont{Sharapov}},
  \bibinfo{journal}{Phys. Rev. B.} \textbf{\bibinfo{volume}{71}},
  \bibinfo{pages}{125124} (\bibinfo{year}{2005}).

\bibitem[{\citenamefont{Miller et~al.}(2009)\citenamefont{Miller, Kubista,
  Rutter, Ruan, de~Heer, First, and Stroscio}}]{Miller2009}
\bibinfo{author}{\bibfnamefont{D.~L.} \bibnamefont{Miller}},
  \bibinfo{author}{\bibfnamefont{K.~D.} \bibnamefont{Kubista}},
  \bibinfo{author}{\bibfnamefont{G.~M.} \bibnamefont{Rutter}},
  \bibinfo{author}{\bibfnamefont{M.}~\bibnamefont{Ruan}},
  \bibinfo{author}{\bibfnamefont{W.~A.} \bibnamefont{de~Heer}},
  \bibinfo{author}{\bibfnamefont{P.~N.} \bibnamefont{First}}, \bibnamefont{and}
  \bibinfo{author}{\bibfnamefont{J.~A.} \bibnamefont{Stroscio}},
  \bibinfo{journal}{Science} \textbf{\bibinfo{volume}{324}},
  \bibinfo{pages}{924} (\bibinfo{year}{2009}).

\bibitem[{\citenamefont{Deacon et~al.}(2007)\citenamefont{Deacon, Chuang,
  Nicholas, Novoselov, and Geim}}]{Deacon2007}
\bibinfo{author}{\bibfnamefont{R.~S.} \bibnamefont{Deacon}},
  \bibinfo{author}{\bibfnamefont{K.-C.} \bibnamefont{Chuang}},
  \bibinfo{author}{\bibfnamefont{R.~J.} \bibnamefont{Nicholas}},
  \bibinfo{author}{\bibfnamefont{K.~S.} \bibnamefont{Novoselov}},
  \bibnamefont{and} \bibinfo{author}{\bibfnamefont{A.~K.} \bibnamefont{Geim}},
  \bibinfo{journal}{Phys. Rev. B.} \textbf{\bibinfo{volume}{76}},
  \bibinfo{pages}{081406} (\bibinfo{year}{2007}), ISSN
  \bibinfo{issn}{1098-0121}.

\bibitem[{\citenamefont{Leadley et~al.}(1989)\citenamefont{Leadley, Nicholas,
  Harris, and Foxon}}]{Leadley1989}
\bibinfo{author}{\bibfnamefont{D.~R.} \bibnamefont{Leadley}},
  \bibinfo{author}{\bibfnamefont{R.~J.} \bibnamefont{Nicholas}},
  \bibinfo{author}{\bibfnamefont{J.~J.} \bibnamefont{Harris}},
  \bibnamefont{and} \bibinfo{author}{\bibfnamefont{C.~T.} \bibnamefont{Foxon}},
  \bibinfo{journal}{Semicond. Sci. Technol.} \textbf{\bibinfo{volume}{4}},
  \bibinfo{pages}{879} (\bibinfo{year}{1989}).

\bibitem[{\citenamefont{Leturcq et~al.}(2003)\citenamefont{Leturcq, L'H\^{o}te,
  Tourbot, Senz, Gennser, Ihn, Ensslin, Dehlinger, and
  Gr\"{u}tzmacher}}]{Leturcq2003}
\bibinfo{author}{\bibfnamefont{R.}~\bibnamefont{Leturcq}},
  \bibinfo{author}{\bibfnamefont{D.}~\bibnamefont{L'H\^{o}te}},
  \bibinfo{author}{\bibfnamefont{R.}~\bibnamefont{Tourbot}},
  \bibinfo{author}{\bibfnamefont{V.}~\bibnamefont{Senz}},
  \bibinfo{author}{\bibfnamefont{U.}~\bibnamefont{Gennser}},
  \bibinfo{author}{\bibfnamefont{T.}~\bibnamefont{Ihn}},
  \bibinfo{author}{\bibfnamefont{K.}~\bibnamefont{Ensslin}},
  \bibinfo{author}{\bibfnamefont{G.}~\bibnamefont{Dehlinger}},
  \bibnamefont{and}
  \bibinfo{author}{\bibfnamefont{D.}~\bibnamefont{Gr\"{u}tzmacher}},
  \bibinfo{journal}{Europhys. Lett.} \textbf{\bibinfo{volume}{61}},
  \bibinfo{pages}{499} (\bibinfo{year}{2003}).

\bibitem[{\citenamefont{Fletcher et~al.}(1992)\citenamefont{Fletcher, Harris,
  Foxon, and Stoner}}]{Fletcher1992}
\bibinfo{author}{\bibfnamefont{R.}~\bibnamefont{Fletcher}},
  \bibinfo{author}{\bibfnamefont{J.~J.} \bibnamefont{Harris}},
  \bibinfo{author}{\bibfnamefont{C.~T.} \bibnamefont{Foxon}}, \bibnamefont{and}
  \bibinfo{author}{\bibfnamefont{R.}~\bibnamefont{Stoner}},
  \bibinfo{journal}{Phys. Rev. B.} \textbf{\bibinfo{volume}{45}},
  \bibinfo{pages}{6659} (\bibinfo{year}{1992}).

\bibitem[{\citenamefont{Gershenson et~al.}(2000)\citenamefont{Gershenson,
  Khavin, Reuter, Schafmeister, and Wieck}}]{Gershenson2000}
\bibinfo{author}{\bibfnamefont{M.}~\bibnamefont{Gershenson}},
  \bibinfo{author}{\bibfnamefont{Y.}~\bibnamefont{Khavin}},
  \bibinfo{author}{\bibfnamefont{D.}~\bibnamefont{Reuter}},
  \bibinfo{author}{\bibfnamefont{P.}~\bibnamefont{Schafmeister}},
  \bibnamefont{and} \bibinfo{author}{\bibfnamefont{A.}~\bibnamefont{Wieck}},
  \bibinfo{journal}{Phys. Rev. Lett.} \textbf{\bibinfo{volume}{85}},
  \bibinfo{pages}{1718} (\bibinfo{year}{2000}).

\bibitem[{\citenamefont{St\"{o}ger et~al.}(1994)\citenamefont{St\"{o}ger,
  Brunthaler, Bauer, Ismail, Meyerson, Lutz, and Kuchar}}]{Stoger1994b}
\bibinfo{author}{\bibfnamefont{G.}~\bibnamefont{St\"{o}ger}},
  \bibinfo{author}{\bibfnamefont{G.}~\bibnamefont{Brunthaler}},
  \bibinfo{author}{\bibfnamefont{G.}~\bibnamefont{Bauer}},
  \bibinfo{author}{\bibfnamefont{K.}~\bibnamefont{Ismail}},
  \bibinfo{author}{\bibfnamefont{B.~S.} \bibnamefont{Meyerson}},
  \bibinfo{author}{\bibfnamefont{J.}~\bibnamefont{Lutz}}, \bibnamefont{and}
  \bibinfo{author}{\bibfnamefont{F.}~\bibnamefont{Kuchar}},
  \bibinfo{journal}{Semicond. Sci. Technol.} \textbf{\bibinfo{volume}{9}},
  \bibinfo{pages}{765} (\bibinfo{year}{1994}).

\bibitem[{\citenamefont{Cao et~al.}(2010)\citenamefont{Cao, Yu, Jauregui, Tian,
  Wu, Liu, Jalilian, Benjamin, Jiang, Bao et~al.}}]{Cao2010}
\bibinfo{author}{\bibfnamefont{H.}~\bibnamefont{Cao}},
  \bibinfo{author}{\bibfnamefont{Q.}~\bibnamefont{Yu}},
  \bibinfo{author}{\bibfnamefont{L.~A.} \bibnamefont{Jauregui}},
  \bibinfo{author}{\bibfnamefont{J.}~\bibnamefont{Tian}},
  \bibinfo{author}{\bibfnamefont{W.}~\bibnamefont{Wu}},
  \bibinfo{author}{\bibfnamefont{Z.}~\bibnamefont{Liu}},
  \bibinfo{author}{\bibfnamefont{R.}~\bibnamefont{Jalilian}},
  \bibinfo{author}{\bibfnamefont{D.~K.} \bibnamefont{Benjamin}},
  \bibinfo{author}{\bibfnamefont{Z.}~\bibnamefont{Jiang}},
  \bibinfo{author}{\bibfnamefont{J.}~\bibnamefont{Bao}}, \bibnamefont{et~al.},
  \bibinfo{journal}{Appl. Phys. Lett.} \textbf{\bibinfo{volume}{96}},
  \bibinfo{pages}{122106} (\bibinfo{year}{2010}).

\bibitem[{\citenamefont{Lara-Avila
  et~al.}(2011{\natexlab{b}})\citenamefont{Lara-Avila, Tzalenchuk, Kubatkin,
  Yakimova, Janssen, Cedergren, Bergsten, and Fal'ko}}]{Lara2011b}
\bibinfo{author}{\bibfnamefont{S.}~\bibnamefont{Lara-Avila}},
  \bibinfo{author}{\bibfnamefont{A.}~\bibnamefont{Tzalenchuk}},
  \bibinfo{author}{\bibfnamefont{S.}~\bibnamefont{Kubatkin}},
  \bibinfo{author}{\bibfnamefont{R.}~\bibnamefont{Yakimova}},
  \bibinfo{author}{\bibfnamefont{T.~J. B.~M.} \bibnamefont{Janssen}},
  \bibinfo{author}{\bibfnamefont{K.}~\bibnamefont{Cedergren}},
  \bibinfo{author}{\bibfnamefont{T.}~\bibnamefont{Bergsten}}, \bibnamefont{and}
  \bibinfo{author}{\bibfnamefont{V.}~\bibnamefont{Fal'ko}},
  \bibinfo{journal}{Phys. Rev. Lett.} \textbf{\bibinfo{volume}{107}},
  \bibinfo{pages}{166602} (\bibinfo{year}{2011}{\natexlab{b}}).

\bibitem[{\citenamefont{Baker et~al.}(2012{\natexlab{b}})\citenamefont{Baker,
  Alexander-Webber, Altebaeumer, Janssen, Tzalenchuk, Lara-Avila, Kubatkin,
  Yakimova, Lin, Li et~al.}}]{Baker2012b}
\bibinfo{author}{\bibfnamefont{A.~M.~R.} \bibnamefont{Baker}},
  \bibinfo{author}{\bibfnamefont{J.~A.} \bibnamefont{Alexander-Webber}},
  \bibinfo{author}{\bibfnamefont{T.}~\bibnamefont{Altebaeumer}},
  \bibinfo{author}{\bibfnamefont{T.~J. B.~M.} \bibnamefont{Janssen}},
  \bibinfo{author}{\bibfnamefont{A.}~\bibnamefont{Tzalenchuk}},
  \bibinfo{author}{\bibfnamefont{S.}~\bibnamefont{Lara-Avila}},
  \bibinfo{author}{\bibfnamefont{S.}~\bibnamefont{Kubatkin}},
  \bibinfo{author}{\bibfnamefont{R.}~\bibnamefont{Yakimova}},
  \bibinfo{author}{\bibfnamefont{C.-T.} \bibnamefont{Lin}},
  \bibinfo{author}{\bibfnamefont{L.-J.} \bibnamefont{Li}},
  \bibnamefont{et~al.}, \bibinfo{journal}{Phys. Rev. B} p.
  \bibinfo{pages}{BW1789} (\bibinfo{year}{2012}{\natexlab{b}}).

\bibitem[{\citenamefont{Lee et~al.}(2010)\citenamefont{Lee, Wijesinghe,
  Diaz-Pinto, and Peng}}]{Lee2010}
\bibinfo{author}{\bibfnamefont{S.}~\bibnamefont{Lee}},
  \bibinfo{author}{\bibfnamefont{N.}~\bibnamefont{Wijesinghe}},
  \bibinfo{author}{\bibfnamefont{C.}~\bibnamefont{Diaz-Pinto}},
  \bibnamefont{and} \bibinfo{author}{\bibfnamefont{H.}~\bibnamefont{Peng}},
  \bibinfo{journal}{Phys. Rev. B.} \textbf{\bibinfo{volume}{82}},
  \bibinfo{pages}{045411} (\bibinfo{year}{2010}).

\bibitem[{\citenamefont{Ki et~al.}(2008)\citenamefont{Ki, Jeong, Choi, Lee, and
  Park}}]{Ki2008}
\bibinfo{author}{\bibfnamefont{D.~K.} \bibnamefont{Ki}},
  \bibinfo{author}{\bibfnamefont{D.}~\bibnamefont{Jeong}},
  \bibinfo{author}{\bibfnamefont{J.~H.} \bibnamefont{Choi}},
  \bibinfo{author}{\bibfnamefont{H.~J.} \bibnamefont{Lee}}, \bibnamefont{and}
  \bibinfo{author}{\bibfnamefont{K.~S.} \bibnamefont{Park}},
  \bibinfo{journal}{Phys. Rev. B.} \textbf{\bibinfo{volume}{78}},
  \bibinfo{pages}{125409} (\bibinfo{year}{2008}).

\bibitem[{\citenamefont{Morozov et~al.}(2006)\citenamefont{Morozov, Novoselov,
  Katsnelson, Schedin, Ponomarenko, Jiang, and Geim}}]{morozov2006}
\bibinfo{author}{\bibfnamefont{S.~V.} \bibnamefont{Morozov}},
  \bibinfo{author}{\bibfnamefont{K.~S.} \bibnamefont{Novoselov}},
  \bibinfo{author}{\bibfnamefont{M.~I.} \bibnamefont{Katsnelson}},
  \bibinfo{author}{\bibfnamefont{F.}~\bibnamefont{Schedin}},
  \bibinfo{author}{\bibfnamefont{L.~A.} \bibnamefont{Ponomarenko}},
  \bibinfo{author}{\bibfnamefont{D.}~\bibnamefont{Jiang}}, \bibnamefont{and}
  \bibinfo{author}{\bibfnamefont{A.~K.} \bibnamefont{Geim}},
  \bibinfo{journal}{Phys. Rev. Lett.} \textbf{\bibinfo{volume}{97}},
  \bibinfo{pages}{016801} (\bibinfo{year}{2006}).

\bibitem[{\citenamefont{Efetov and Kim}(2010)}]{EfKim2010}
\bibinfo{author}{\bibfnamefont{D.~K.} \bibnamefont{Efetov}} \bibnamefont{and}
  \bibinfo{author}{\bibfnamefont{P.}~\bibnamefont{Kim}},
  \bibinfo{journal}{Phys. Rev. Lett.} \textbf{\bibinfo{volume}{105}},
  \bibinfo{pages}{256805} (\bibinfo{year}{2010}).

\bibitem[{\citenamefont{Chen et~al.}(2008)\citenamefont{Chen, Jang, Xiao,
  Ishigami, and Fuhrer}}]{newChen08}
\bibinfo{author}{\bibfnamefont{J.-H.} \bibnamefont{Chen}},
  \bibinfo{author}{\bibfnamefont{C.}~\bibnamefont{Jang}},
  \bibinfo{author}{\bibfnamefont{S.}~\bibnamefont{Xiao}},
  \bibinfo{author}{\bibfnamefont{M.}~\bibnamefont{Ishigami}}, \bibnamefont{and}
  \bibinfo{author}{\bibfnamefont{M.}~\bibnamefont{Fuhrer}},
  \bibinfo{journal}{Nature Nanotechnology} \textbf{\bibinfo{volume}{3}},
  \bibinfo{pages}{206} (\bibinfo{year}{2008}).

\bibitem[{\citenamefont{Song et~al.}(2012)\citenamefont{Song, Reizer, and
  Levitov}}]{Song2012}
\bibinfo{author}{\bibfnamefont{J.~C.~W.} \bibnamefont{Song}},
  \bibinfo{author}{\bibfnamefont{M.~Y.} \bibnamefont{Reizer}},
  \bibnamefont{and} \bibinfo{author}{\bibfnamefont{L.~S.}
  \bibnamefont{Levitov}}, \bibinfo{journal}{Phys. Rev. Lett.}
  \textbf{\bibinfo{volume}{109}}, \bibinfo{pages}{106602}
  (\bibinfo{year}{2012}).

\bibitem[{\citenamefont{Graham et~al.}(2013)\citenamefont{Graham, Shi, Ralph,
  Park, and McEuen}}]{Graham2013}
\bibinfo{author}{\bibfnamefont{M.~W.} \bibnamefont{Graham}},
  \bibinfo{author}{\bibfnamefont{S.-F.} \bibnamefont{Shi}},
  \bibinfo{author}{\bibfnamefont{D.~C.} \bibnamefont{Ralph}},
  \bibinfo{author}{\bibfnamefont{J.}~\bibnamefont{Park}}, \bibnamefont{and}
  \bibinfo{author}{\bibfnamefont{P.~L.} \bibnamefont{McEuen}},
  \bibinfo{journal}{Nature Physics} \textbf{\bibinfo{volume}{9}},
  \bibinfo{pages}{10.1038/nphys2493} (\bibinfo{year}{2013}).

\bibitem[{\citenamefont{Betz et~al.}(2013)\citenamefont{Betz, Jhang, Pallecchi,
  Ferreira, Feve, Berroir, and Placais}}]{Betz2013}
\bibinfo{author}{\bibfnamefont{A.}~\bibnamefont{Betz}},
  \bibinfo{author}{\bibfnamefont{S.~H.} \bibnamefont{Jhang}},
  \bibinfo{author}{\bibfnamefont{E.}~\bibnamefont{Pallecchi}},
  \bibinfo{author}{\bibfnamefont{R.}~\bibnamefont{Ferreira}},
  \bibinfo{author}{\bibfnamefont{G.}~\bibnamefont{Feve}},
  \bibinfo{author}{\bibfnamefont{J.-M.} \bibnamefont{Berroir}},
  \bibnamefont{and} \bibinfo{author}{\bibfnamefont{B.}~\bibnamefont{Placais}},
  \bibinfo{journal}{Nature Physics} \textbf{\bibinfo{volume}{9}},
  \bibinfo{pages}{10.1038/nphys2494} (\bibinfo{year}{2013}).

\end{thebibliography}


\end{document}